%% file: main.tex
\documentclass[letterpaper]{article} 
\usepackage[submission]{aaai2026}  
\usepackage{times}  
\usepackage{helvet}  
\usepackage{courier}  
\usepackage[hyphens]{url}  
\usepackage{graphicx} 
\usepackage{natbib}  
\usepackage{caption} 
\usepackage{algorithm}
\usepackage{algorithmic}

\usepackage{newfloat}
\usepackage{listings}
\DeclareCaptionStyle{ruled}{labelfont=normalfont,labelsep=colon,strut=off} 
\floatstyle{ruled}
\newfloat{listing}{tb}{lst}{}
\floatname{listing}{Listing}
\makeatletter
\def\showauthors@on{T}
\makeatother

\title{Toward Automated and Trustworthy Scientific Analysis and Visualization with LLM-Generated Code}
\author{
    Apu Kumar Chakroborti \textsuperscript{\rm 1},
    Yi Ding \textsuperscript{\rm 1},
    Lipeng Wan\textsuperscript{\rm 1}
}
\affiliations{
    \textsuperscript{\rm 1}Georgia State University\\
    achakroborti1@student.gsu.edu, yiding@gsu.edu, lwan@gsu.edu
}

\usepackage{bibentry}

\begin{document}

\maketitle



\input{0.abstract}






\input{1.introduction}

\input{2.background}

\input{3.methodology}

\input{4.evaluation}

\input{5.related_work}
\input{6.conclusion}
\section{Acknowledgments}
Research was sponsored by the Army Research Laboratory and was accomplished under Cooperative Agreement Number W911NF-23-2-0224. The views and conclusions contained in this document are those of the authors and should not be interpreted as representing the official policies, either expressed or implied, of the Army Research Laboratory or the U.S. Government. The U.S. Government is authorized to reproduce and distribute reprints for Government purposes notwithstanding any copyright notation herein.



\bibliography{references}

\end{document}

%% file: 0.abstract.tex
\begin{abstract}
    
As modern science becomes increasingly data-intensive, the ability to analyze and visualize large-scale, complex datasets is critical to accelerating discovery. However, many domain scientists lack the programming expertise required to develop custom data analysis workflows, creating barriers to timely and effective insight. Large language models (LLMs) offer a promising solution by generating executable code from natural language descriptions. In this paper, we investigate the trustworthiness of open-source LLMs in autonomously producing Python scripts for scientific data analysis and visualization. We construct a benchmark suite of domain-inspired prompts that reflect real-world research tasks and systematically evaluate the executability and correctness of the generated code. Our findings show that, without human intervention, the reliability of LLM-generated code is limited, with frequent failures caused by ambiguous prompts and the models’ insufficient understanding of domain-specific contexts. To address these challenges, we design and assess three complementary strategies: data-aware prompt disambiguation, retrieval-augmented prompt enhancement, and iterative error repair. While these methods significantly improve execution success rates and output quality, further refinement is needed. This work highlights both the promise and current limitations of LLM-driven automation in scientific workflows and introduces actionable techniques and a reusable benchmark for building more inclusive, accessible, and trustworthy AI-assisted research tools.

\end{abstract}

%% file: 1.introduction.tex
\section{Introduction}

In modern science, the integration of advanced technologies and instruments has led to an unprecedented surge in data generation across disciplines such as physics, chemistry, biology, and environmental science~\cite{gray2005scientific}. The ability to analyze and interpret this data is essential for driving scientific discovery~\cite{kehrer2012visualization}. Traditionally, this analysis has relied on domain scientists manually writing code to extract insights from complex datasets. While effective, this manual approach is often labor-intensive~\cite{fuchs2009visualization}, error-prone, and increasingly infeasible given the scale and complexity of today’s data~\cite{steed2013big}.

Large language models (LLMs) offer a promising alternative by automating the generation of code for scientific data analysis and visualization~\cite{nejjar2025llms}. Models like GPT-4~\cite{achiam2023gpt} and Claude 3.5~\cite{electronics13132657} exhibit impressive capabilities in natural language understanding and code generation. Trained on large-scale text and code corpora, they can translate natural language queries into executable Python scripts~\cite{ridnik2024codegenerationalphacodiumprompt, zhang2024codeagentenhancingcodegeneration}, enabling a new class of intelligent tools for scientific workflows. This shift has the potential to accelerate research by lowering the entry barrier to complex analysis tasks.

A major motivation for integrating LLMs into scientific analysis is the technical barrier faced by domain experts. Many researchers possess deep scientific knowledge but lack proficiency in programming or experience with data processing libraries and visualization frameworks~\cite{springmeyer1992characterization}. This challenge is especially pronounced in disciplines where computational training is not standard, such as biology, medicine, or environmental science. By automating code generation, LLMs allow scientists to focus on research questions rather than implementation details. For example, a researcher studying climate trends can generate visualization code for multidimensional geospatial data without manually writing boilerplate scripts. This capability empowers broader participation in data-driven science and facilitates rapid experimentation with advanced analytical methods.

Despite these benefits, LLM-generated code is not without limitations. A core challenge is reliability. While LLMs are proficient at interpreting general instructions, they often struggle with domain-specific nuances. The code they produce may include logical flaws, inappropriate library usage, or incorrect assumptions about data structures~\cite{10449667}. These issues are exacerbated when users provide vague prompts or lack a clear understanding of the dataset. As a result, generated scripts may fail to execute or produce misleading results. Diagnosing such failures often requires significant programming knowledge, undermining the accessibility LLMs aim to provide. This trial-and-error cycle can slow scientific progress and, in some cases, compromise the integrity of results. An obvious remedy might be to fine-tune or retrain LLMs on domain-specific datasets. However, retraining is resource-intensive, requiring extensive computational infrastructure and large volumes of curated training data. Moreover, domain-specialized models risk reduced generalization, increased maintenance burdens, and potential overfitting. Given these trade-offs, lightweight, model-agnostic strategies are more preferable. 

In this work, we evaluate the trustworthiness of open-source LLMs in autonomously generating code for scientific data analysis and visualization, with the goal of minimizing the need for manual coding in routine research workflows. We simulate natural language prompts that reflect how domain experts typically describe tasks involving diverse datasets, domain-specific analytical methods, and complex visualizations. These prompts are submitted to LLMs, which generate Python scripts that we evaluate by executing and assessing for correctness. Our results show that the success rate of LLM-generated code without human intervention remains low. We then implement and asses three model-agnostic strategies aimed at improving the reliability of generated code and aligning it more closely with user intent and dataset structure: (1) \textbf{Data-Aware Prompt Disambiguation} enriches prompts with schema-level metadata and sample statistics to reduce ambiguity and guide the model toward more accurate code generation. (2) \textbf{Retrieval-Augmented Prompt Enhancement} supplements prompts with relevant examples from domain-specific code repositories to ground the model’s output in realistic and contextually appropriate workflows. (3) \textbf{Iterative Error Repair} enables the model to revise its own code based on runtime or compilation errors, mimicking a lightweight debugging loop. While our results show that these techniques significantly increase the likelihood that the generated code executes successfully and produces meaningful visualizations, further improvement is still needed.

The contributions of this paper are as follows:
\begin{itemize}
    \item We conduct a systematic evaluation of state-of-the-art open-source LLMs in the context of scientific data analysis and visualization, focusing on their ability to autonomously generate executable Python code from natural language prompts reflective of domain expert intent.
    \item We design a benchmark suite of synthetic prompts modeled after real-world use cases in diverse scientific domains, enabling a controlled and reproducible assessment of LLM-generated code reliability and correctness.
    \item We implement and evaluate three strategies to improve code reliability. Our results demonstrate that while these strategies substantially increase the success rate of LLM-generated code, challenges remain.
\end{itemize}

%% file: 2.background.tex
\section{Background}

\subsection{Scientific Data Analysis and Visualization}
Scientific data analysis and visualization are essential across domains such as physics, biology, climate science, and engineering, enabling researchers to extract insights from vast amounts of data generated by experiments, simulations, and observations~\cite{springmeyer1992characterization}. Unlike general-purpose datasets, scientific data is often high-dimensional, structured, and stored in specialized formats such as HDF5~\cite{10.1145/1966895.1966900}, NetCDF~\cite{56302}, and FITS~\cite{pence2010definition}. These formats support hierarchical organization and rich metadata, but also increase complexity when parsing, interpreting, and visualizing data programmatically.

Scientific datasets typically require specialized preprocessing steps—such as coordinate transformations, anomaly detection, or unit conversions—tailored to the specific domain. Visualization techniques must also be domain-aware, varying from volumetric rendering in medical imaging to contour plots in fluid dynamics. As a result, domain scientists often rely on tools like NumPy, SciPy, Matplotlib, and VTK~\cite{vtkBook}. However, developing and debugging such code can be time-consuming and error-prone, particularly for those without strong programming backgrounds.

\subsection{LLM-Based Code Generation}

Large language models (LLMs) have emerged as powerful tools for automated code generation~\cite{jiang2024surveylargelanguagemodels}, capable of producing executable scripts from natural language prompts~\cite{ridnik2024codegenerationalphacodiumprompt}. These models support a wide range of programming tasks—including code completion~\cite{husein2024large}, debugging~\cite{tian2024debugbench}, optimization~\cite{gao2024search}, and documentation~\cite{sun2024source}—and are increasingly integrated into modern development workflows.

In scientific computing, LLMs offer an opportunity to accelerate research and lower barriers for non-programmers by translating high-level analytical goals into executable code~\cite{hong2024data}. This is especially impactful in fields like genomics and climate science, where datasets are massive, complex, and stored in structured formats. However, existing models face key limitations. LLMs lack true contextual awareness of data and execution environments, making them prone to errors when handling multidimensional arrays or hierarchical file structures~\cite{tony2025promptingtechniquessecurecode}. Ambiguous prompts often lead to code that is inefficient, incorrect, or fails to execute.
Moreover, debugging LLM-generated code remains difficult for non-expert users, who may not be equipped to interpret runtime errors or identify logical flaws~\cite{sarker2024syntacticrobustnessllmbasedcode}. These challenges highlight the need for strategies that bridge the gap between user intent, data context, and reliable code generation—especially in high-stakes domains like scientific research.

%% file: 3.methodology.tex
\section{Methodology} 

\subsection{Data Collection and Preparation}

To evaluate LLM-based code generation in scientific data analysis, we curated a benchmark composed of diverse hierarchical datasets from three primary sources: the MatPlotBench dataset~\cite{yang2024matplotagentmethodevaluationllmbased}, datasets from NASA’s Earth Observing System (EOS) ~\cite{hdf_eos_tools_and_information_center}, and medical imaging datasets from NYU Langone Health’s fastMRI repository~\cite{zbontar2019fastmriopendatasetbenchmarks}. These datasets were selected for their structured formats and relevance to real-world scientific workflows.

We first prepared 12 datasets from MatPlotBench, originally in CSV format and accompanied by natural language descriptions of expected visualizations. To enable consistent evaluation across data types, we converted these CSV files to HDF5 format, preserving column names as dataset identifiers. While these datasets lack deep hierarchies, they represent common scientific tabular data used in visualization tasks. Next, we collected 61 HDF5 datasets from NASA’s EOS repositories via the HDF-EOS Tools and Information Center~\cite{773470}. These files contain multi-level hierarchies and embedded metadata, typical of remote sensing and environmental monitoring data. We also gathered associated Python scripts to guide the construction of prompts grounded in domain-relevant processing and visualization logic. Finally, we curated 11 MRI datasets from NYU's fastMRI repository, originally stored in HDF5 and DICOM formats. To standardize the format, we converted DICOM files to HDF5, preserving key metadata and ensuring compatibility with our benchmark. Accompanying visualization scripts were used to develop prompts that capture domain-specific imaging tasks.

This benchmark spans a range of structural complexities, from flat tabular data to deeply nested hierarchies, allowing us to systematically assess LLMs' code generation capability across different scientific data modalities.

\subsection{Prompt Generation}
To evaluate the ability of LLMs to generate scientific analysis and visualization code, we manually constructed two types of natural language prompts for each dataset: \emph{detailed} and \emph{simplified}. These prompts were derived from existing Python scripts ~\cite{Neural_Data_Science_in_Python} and designed to reflect how users with varying levels of technical specificity might request code generation in real-world scenarios.

We began by analyzing the original scripts to identify key operations such as data loading, preprocessing, transformation, and visualization. Based on this understanding, we created detailed prompts that included explicit instructions, such as paths of HDF5 datasets, library usage (e.g., h5py, NumPy, Matplotlib), and the desired visualization format. In contrast, simplified prompts provided a more general description of the task, omitting technical specifics and simulating how a non-expert or high-level user might formulate the same request. For example, a detailed prompt might state: ``Generate Python code that loads an HDF5 file using h5py, accesses the dataset at \emph{/temperature/data}, converts it to a NumPy array, and plots it as a time-series line graph using Matplotlib.'' The corresponding simplified prompt would be: ``Generate a Python script to extract and visualize temperature data from an HDF5 file.''

This two-tiered prompting strategy enabled us to assess LLM performance under different levels of prompt specificity and better understand their robustness and adaptability to realistic scientific computing queries.

\subsection{Data-Aware Prompt Disambiguation}
Scientific datasets stored in HDF5 format are typically organized in hierarchical structures, where each dataset or attribute is accessed via a full path (e.g., \emph{/group1/subgroup2/temperature}). However, users, especially those unfamiliar with the dataset’s structure, often reference variables using vague or abbreviated terms. For instance, a user might request to ``visualize the temperature data'' without specifying whether the dataset resides at \emph{/measurements/temperature}, or another location. Such ambiguity frequently causes LLMs to generate incorrect or non-functional code, as they depend solely on the prompt text and lack direct access to the data hierarchy.

To address this challenge, we evaluate a data-aware prompt disambiguation approach that augments user queries with accurate structural references extracted directly from the HDF5 file. Rather than modifying the user’s intent, our method enriches the prompt with minimal, contextually relevant metadata, such as precise dataset paths and attribute names, allowing LLMs to generate semantically correct and executable Python code. This strategy preserves the natural language character of the prompt while grounding it in the actual data structure, helping the LLM resolve ambiguity and avoid hallucinated references.

The disambiguation process begins by parsing the HDF5 file to enumerate all dataset paths and their associated attributes. The original user prompt is then tokenized into monograms and bigrams by stripping special characters and splitting on whitespace, punctuation, and slashes. These tokens are matched against the dataset paths using a multi-stage process. Initially, the system attempts exact and partial string matches, ignoring formatting differences such as punctuation. If no matches are found, fuzzy matching is applied using the Levenshtein distance via the RapidFuzz library. Two thresholds are used: a strict threshold (87\%) followed by a relaxed threshold (80\%) to tolerate greater variation. Tokenization is extended by splitting long terms using slashes to improve matching for nested dataset paths. The prompt is also reprocessed to regenerate bigrams, helping identify compound dataset names that may not be captured by single-token comparisons.

Using the cleaned and expanded tokens, the system performs a second round of dataset matching based on four criteria: exact full-path match, subgroup-level match, partial name match, and fuzzy similarity. All matching results are collected and ranked, prioritizing full-path and subgroup matches over less precise alternatives. For each matched dataset, relevant attribute names are also retrieved. These matched elements are then appended to the original prompt in a concise and non-intrusive format. The resulting prompt remains intelligible to the user while supplying the LLM with structurally grounded context, significantly improving its ability to generate valid and relevant code.

This method is model-agnostic and requires no changes to the LLM architecture. It improves code generation quality by aligning prompt content with the underlying data structure, thereby reducing the risk of hallucinated references or invalid access paths.

\subsection{Retrieval-Augmented Prompt Enhancement}
While LLMs show impressive proficiency in generating Python code from natural language prompts, their reliability often degrades in domain-specific scientific workflows that require precise, context-aware decisions. Scientific tasks such as accessing hierarchical datasets, preprocessing structured data, and selecting appropriate visualization techniques frequently involve implicit assumptions and domain knowledge that are difficult to express fully in natural language. As a result, LLM-generated scripts can be incomplete, incorrect, or overly generic.

To address these limitations, we adopt and evaluate a retrieval-augmented prompt enhancement strategy, inspired by the contextual programming paradigm introduced in REDCODER~\cite{parvez2021retrieval}. In that setting, programmers curate notes and code snippets while learning a new language to support future problem-solving. Analogously, our approach augments user prompts with relevant task-specific examples drawn from a curated repository of scientific code snippets. This process does not represent a novel technique, but rather an adaptation of prior work to the domain of scientific data analysis and visualization, allowing us to empirically assess its effectiveness.

We constructed a lightweight, domain-specific knowledge base to support retrieval-augmented prompting, organized into three targeted indexes: (1) dataset and attribute access, (2) data preprocessing, and (3) scientific visualization. For embedding generation, we used the pretrained all-MiniLM-L6-v2 model from the Sentence-Transformers library~\cite{reimers2019sentence}, and built the corresponding vector indexes using the Faiss library~\cite{douze2024faiss}. Each indexed entry consists of a concise title, a brief natural language description, and an illustrative Python code snippet. The access index captures common usage patterns for loading structured data from HDF5 files with h5py; the preprocessing index includes examples such as masking invalid values, performing unit conversions, and detecting outliers; and the visualization index provides templates for generating time-series plots, heatmaps, and 3D renderings using tools like Matplotlib and VTK.

When given a user prompt, we first use an LLM to decompose the query into three components corresponding to the index categories: data access intent, preprocessing requirements, and visualization goals. We then retrieve the most semantically relevant entry from each index using similarity-based matching. These examples—comprising both explanatory text and code—are appended to the original prompt, resulting in a retrieval-augmented version that offers clearer context to the LLM.

This technique improves LLM output by grounding it in domain-relevant usage patterns, especially when the original prompt is vague or underspecified. By exposing the model to established idioms and best practices from scientific computing, we help guide code generation toward more syntactically valid and semantically meaningful solutions.

\subsection{Iterative Error Repair}
While LLMs have shown substantial potential in automating code generation, their outputs—especially in scientific computing contexts—often contain syntax errors, incorrect logic, or misused APIs that prevent successful execution. Prior work has suggested using iterative refinement strategies, where models revise their outputs based on execution feedback, as a practical way to improve code correctness. In this study, we adopt and evaluate such an iterative error repair approach to assess its effectiveness in improving the reliability of LLM-generated scientific analysis and visualization scripts.

In our evaluation framework, each generated Python script is executed in a controlled runtime environment, where we capture both the standard output and any error messages. We classify the outcome of each script into three categories: (1) \emph{runnable and correct (simplified as ``correct'')}, where the script runs successfully and produces the expected output; (2) \emph{runnable but incorrect (simplified as ``runnable'')}, where it executes without errors but generates invalid or incomplete results; and (3) \emph{failed to run (simplified as ``failed'')}, where execution is halted due to syntax or runtime errors.

For scripts that fail to execute, we extract the associated error messages and append them to the original prompt. This augmented prompt is then resubmitted to the LLM, which attempts to generate a revised version of the code. The new script is evaluated again, and if it still fails, the process repeats—feeding updated error traces back into the model—until a correct script is produced or a predefined maximum number of repair attempts is reached. This process emulates a lightweight debugging loop and allows the model to adjust its output in response to concrete runtime signals.

Our evaluation focuses on how effectively this iterative strategy helps LLMs recover from failure and produce functional, accurate scripts for complex scientific tasks. We observe that even a few rounds of error-informed refinement can substantially improve execution success rates, especially for tasks involving unfamiliar APIs, hierarchical data access, or incomplete initial prompts. While this technique does not eliminate all failure cases, it provides a low-overhead, model-agnostic mechanism for improving code reliability without requiring retraining or fine-tuning.

%% file: 4.evaluation.tex
\section{Evaluation}
\subsection{Evaluation Testbed}
All experiments were performed on a GPU server. The system is equipped with two NVIDIA RTX A4500 GPUs, each featuring 20GB of GDDR6 memory, enabling high-throughput, parallel processing for low-latency prompt handling and efficient model inference. Complementing the GPU setup, the server is powered by dual Intel Xeon Silver 4210R processors (10 cores and 20 threads per CPU, base clock 2.40GHz, turbo boost up to 3.20 GHz), offering strong multi-threaded performance typical of data center-grade workloads. The machine also includes 64GB of ECC DDR4 memory (8×8GB, 2933MHz), providing reliable, high-capacity memory for managing large prompt contexts, intermediate outputs, and memory-intensive tasks such as iterative error repair. 

To manage and run the LLMs, we utilized Ollama~\cite{ollama2023}, a lightweight and efficient framework for serving and interacting with open-source large language models locally. Ollama provides an optimized runtime environment that minimizes overhead and enhances performance when running large models on local hardware. It allows seamless interaction with different LLMs, facilitating rapid testing and evaluation of various models without requiring access to cloud-based APIs. For our experiments, we selected DeepSeek-R1-70B~\cite{guo2025deepseek}, Gemma3-27B~\cite{team2025gemma}, Llama3-70B~\cite{siriwardhana2024domainadaptationllama370binstructcontinual}, Devstral-24B~\cite{Devstral2025} and Magicoder-7B~\cite{wei2024magicoderempoweringcodegeneration} as our test models. 

\subsection{Impact of Prompt Quality}

Prompt quality plays a critical role in determining the accuracy, reliability, and executability of LLM-generated code. Well-structured prompts that explicitly specify dataset paths, attribute names, function usage, and intended analysis or visualization goals provide essential context for guiding LLMs toward generating valid and semantically meaningful scripts. In contrast, ambiguous or underspecified prompts often result in code that is incorrect, incomplete, or fails to execute, particularly in domains that involve complex data structures like hierarchical HDF5 files.

To systematically assess how prompt quality affects code generation performance, we evaluated each LLM using two prompt variants: detailed and simple. For the MatPlotBench dataset, both prompt types were directly obtained from the benchmark suite. For the NASA's EOS and fastMRI datasets, we manually authored detailed prompts that closely mirrored real-world scientific computing scenarios. These prompts included precise information about file structure, data access paths, required preprocessing operations, and expected visualization outputs. To create simple prompts, we used Llama3-70B to summarize the detailed versions into more concise, natural-language queries. These simplified prompts omitted many technical details, mimicking how a domain scientist with limited programming expertise might describe their analysis intent.

\begin{figure}[h] \centering
\vspace{-2mm}
\includegraphics[width=0.99\columnwidth]{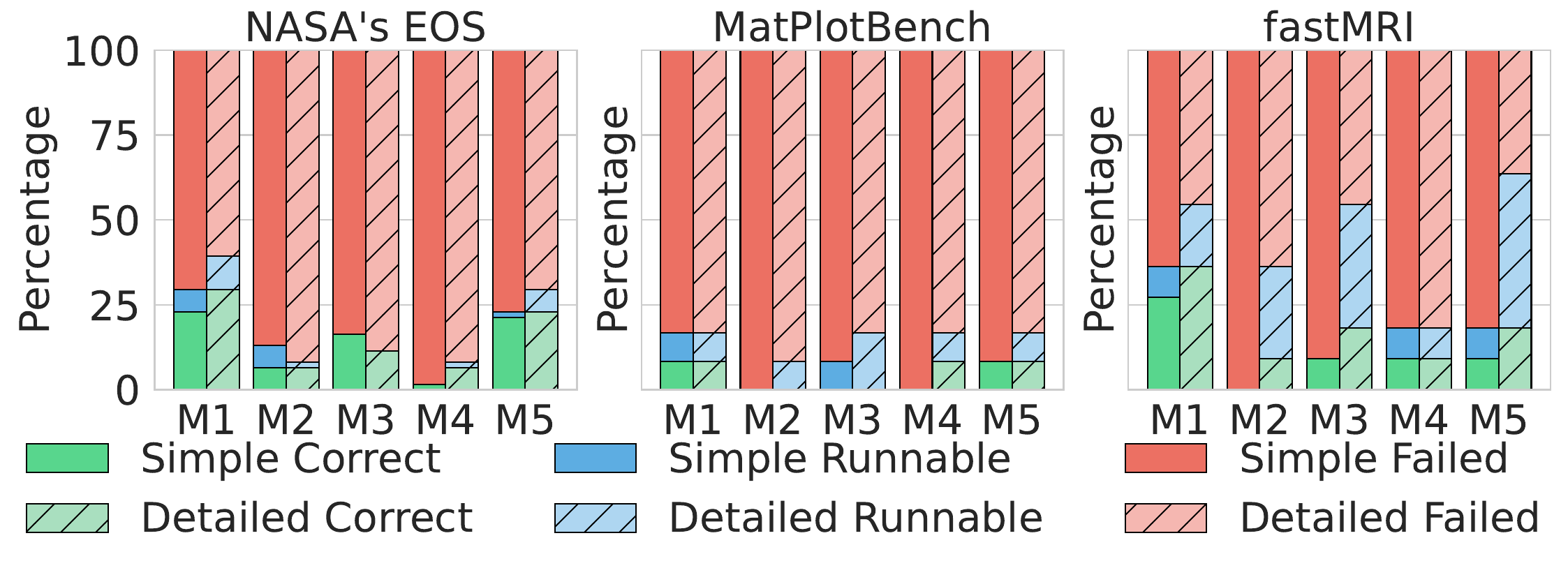}
\vspace{-2mm}
\caption{Execution results of data analysis/visualization codes generated by different LLMs using simple and detailed prompts (M1: Devstral-24B, M2: Magicoder-7B, M3: Llama3-70B, M4: Gemma3-27B, M5: DeepSeek-R1-70B).}
\label{combined_3_datasets_simple_vs_expert}
\vspace{-2mm}
\end{figure}

We then used both prompt types to generate Python scripts with several open-source LLMs and executed the outputs in a controlled environment. The results were categorized into three classes: successful execution with correct output (``correct''), execution with incorrect or partial results (``runnable''), and execution failure (``failed''), and we computed the percentage distribution across these categories. As shown in Figure~\ref{combined_3_datasets_simple_vs_expert}, while most LLMs struggled even with detailed prompts, their performance dropped further when provided with simpler ones. For example, with NASA's EOS datasets, approximately 30\% of the scripts generated by Devstral-24B using detailed prompts executed correctly and produced valid visualizations or analyses. However, this success rate dropped to 21\% with simple prompts. Other models performed worse, with success rates declining across the board when key contextual information was omitted.

These findings underscore the sensitivity of LLMs to prompt quality in scientific computing tasks. Although detailed prompts can improve performance, they often require technical knowledge that end users may not possess. This highlights the importance of automated techniques—such as data-aware disambiguation and retrieval-based augmentation—for enriching user prompts with accurate, task-relevant information, thereby reducing ambiguity and improving the overall reliability of LLM-generated code.

\subsection{Impact of Data-Aware Prompt Disambiguation}
To evaluate the effectiveness of data-aware prompt disambiguation, we conducted experiments using simple prompts and five different open-source LLMs. For each LLM, we compared the quality of the generated Python scripts under two conditions: with and without prompt disambiguation. This technique modifies the input prompt by automatically identifying and replacing ambiguous or loosely specified dataset and attribute references with accurate paths extracted from the underlying HDF5 file. The goal is to clarify how data should be accessed without altering the user's original intent or analysis objective.

\begin{figure}[h] \centering
\vspace{-2mm}
\includegraphics[width=0.99\columnwidth]{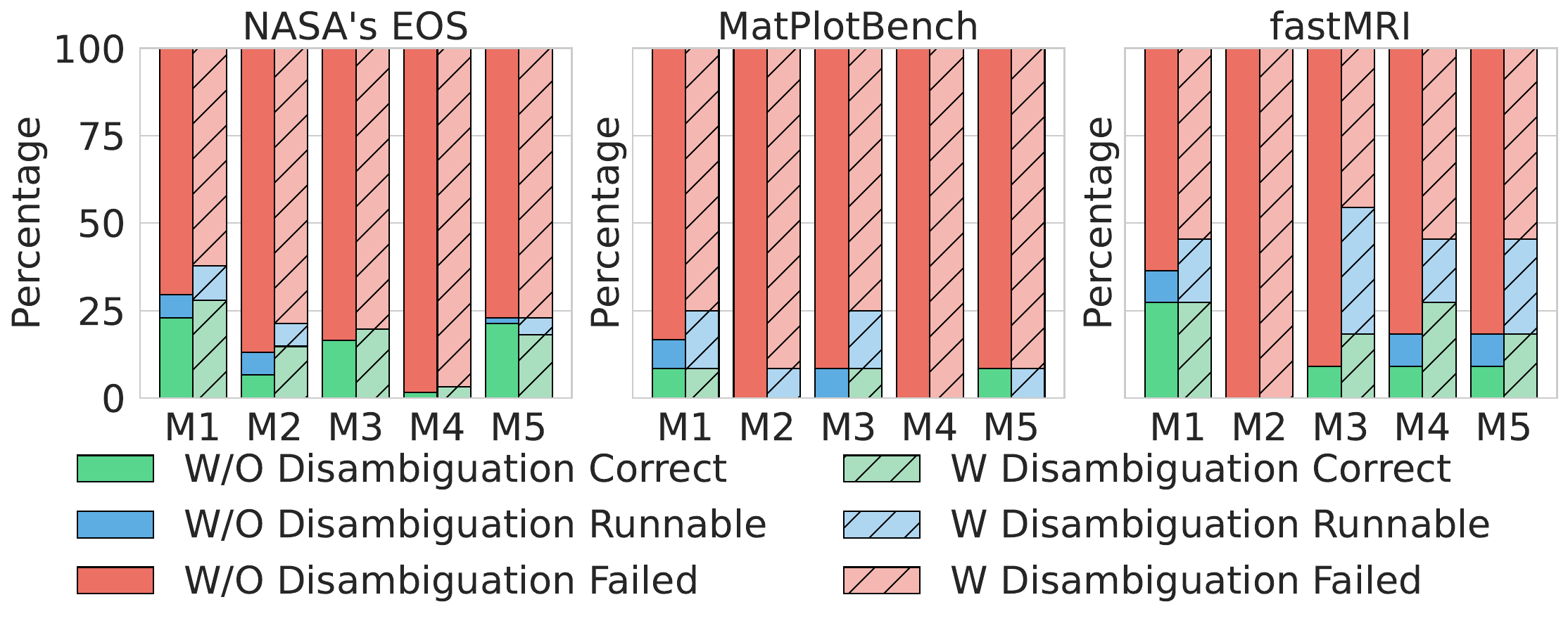}
\vspace{-2mm}
\caption{Execution results of data analysis/visualization codes generated by different LLMs using simple prompts without and with data-aware prompt disambiguation enabled (M1: Devstral-24B, M2: Magicoder-7B, M3: Llama3-70B, M4: Gemma3-27B, M5: DeepSeek-R1-70B).}
\label{simple_query_combined_3_datasets_wc_vs_wo_c}
\vspace{-2mm}
\end{figure}

As shown in Figure~\ref{simple_query_combined_3_datasets_wc_vs_wo_c}, incorporating data-aware prompt disambiguation led to a noticeable increase in the number of scripts that executed without runtime errors across all models. This improvement is especially evident in cases where the original prompt failed to specify exact dataset names or paths, leading LLMs to hallucinate or incorrectly infer data locations. By grounding these references in the actual dataset structure, our method reduced such failures and made the generated code more robust to prompt underspecification.

However, while script executability improved, the proportion of outputs that produced correct analytical or visualization results showed only marginal gains. This is expected, as disambiguation primarily addresses issues related to data access (e.g., locating and loading the correct variables from an HDF5 file), but does not improve the specification of the broader analytical task. In other words, if the original prompt lacked clarity about preprocessing steps, visualization goals, or the appropriate use of analytical functions, those deficiencies persisted even after disambiguation. Consequently, scripts may still execute but fail to produce meaningful or accurate outputs. Nevertheless, this shift is a valuable intermediate improvement. Although incorrect outputs are not ideal, they are generally easier for users to debug and fix compared to runtime failures. 

These findings underscore the importance of separating two key challenges in LLM-assisted scientific computing: (1) resolving structural ambiguities that hinder execution and (2) ensuring semantic clarity that guides correct analysis and visualization. While data-aware prompt disambiguation significantly enhances the former, achieving high-quality end-to-end outcomes requires complementary strategies, such as retrieval-augmented prompt enhancement or iterative refinement, to improve task specification. Together, these techniques can serve as modular components in a robust pipeline for trustworthy LLM-based scientific code generation.

\subsection{Impact of Retrieval-Augmented Prompt Enhancement}

To evaluate the effectiveness of retrieval-augmented prompt enhancement, we conducted a comparative study using the same set of simple prompts as in previous experiments, with data-aware prompt disambiguation enabled by default. This setup isolates the additional impact of retrieval-based augmentation. For each user query, our system extracts sub-intents—such as data access, preprocessing, and visualization—and retrieves relevant code examples and explanations from curated repositories, which are then appended to the prompt. These augmented prompts aim to provide the LLM with more task-specific context without requiring the user to supply detailed instructions.

\begin{figure}[h] \centering
\vspace{-2mm}
\includegraphics[width=0.99\columnwidth]{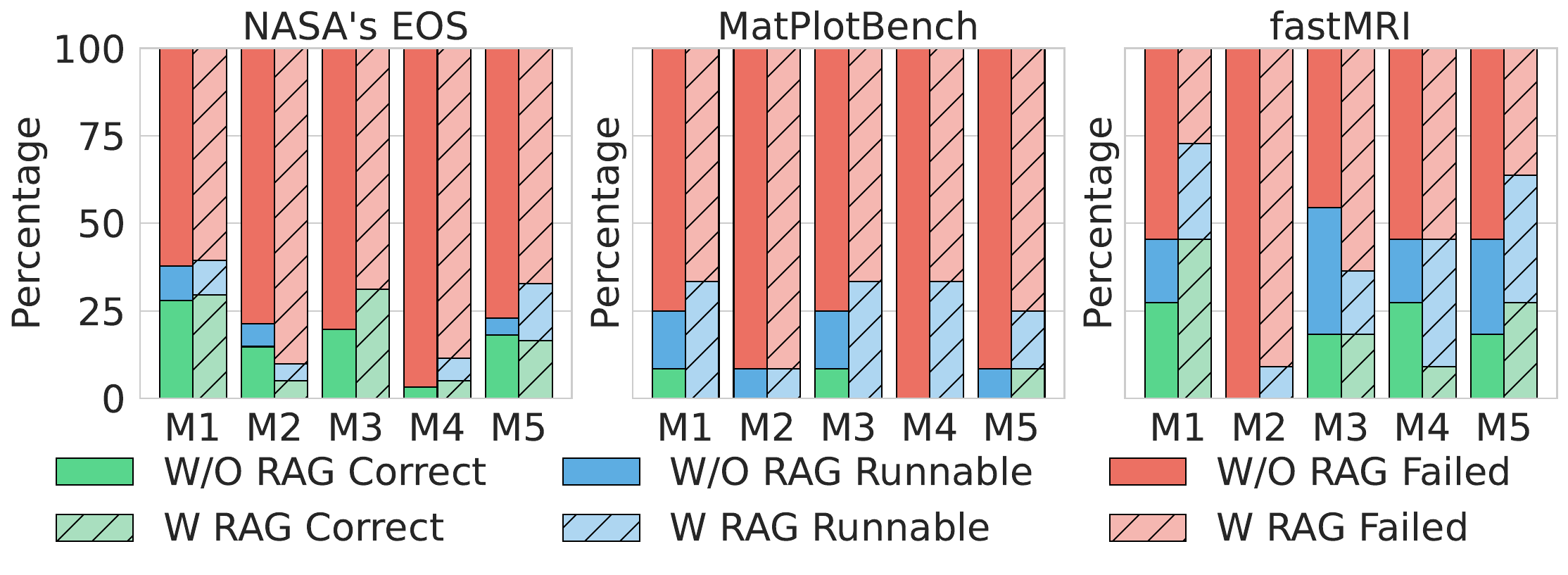}
\vspace{-2mm}
\caption{Execution results of data analysis/visualization codes generated by different LLMs using simple prompts (data-aware prompt disambiguation is enabled by default) without and with retrieval-augmented prompt enhancement enabled (M1: Devstral-24B, M2: Magicoder-7B, M3: Llama3-70B, M4: Gemma3-27B, M5: DeepSeek-R1-70B).}
\label{simple_query_combined_3_datasets_w_rag_vs_wo_rag}
\vspace{-2mm}
\end{figure}

As illustrated in Figure~\ref{simple_query_combined_3_datasets_w_rag_vs_wo_rag}, the improvements enabled by retrieval augmentation vary across models and datasets. For example, on the NASA EOS dataset, enabling retrieval augmentation led to a surprising decline in performance for Magicoder-7B, indicating potential issues with the model’s ability to incorporate retrieved context effectively or a mismatch between the retrieved examples and the dataset’s structural complexity. Conversely, Devstral-24B showed no significant improvement on the same dataset, suggesting that for LLMs already capable of handling metadata-enriched prompts, retrieval augmentation may yield diminishing returns when the task is dominated by structural challenges rather than logic or procedural knowledge.

In contrast, the benefit of retrieval-based augmentation becomes more apparent in domains requiring domain-specific visualizations or preprocessing logic. For instance, on the fastMRI dataset, Devstral-24B exhibited a substantial improvement in the number of correct outputs when retrieval examples were included. This suggests that retrieval augmentation is particularly helpful when the model must learn domain-specific operations—such as masking invalid pixels in MRI scans or adjusting visualization parameters for 3D medical volumes—that are not easily inferred from generic prompts or metadata alone.

These findings reveal that retrieval augmentation is most effective when the retrieved examples closely match the procedural and semantic intent of the user’s query. When alignment is weak, or when models are overwhelmed by additional context, performance can stagnate or even regress. Therefore, the success of this method hinges on both the quality of the retrieved examples and the LLM's capacity to integrate them meaningfully.

\subsection{Impact of Iterative Error Repair}
To assess the effectiveness of iterative error repair, we conducted experiments using simple prompts with the Devstral-24B model. Both data-aware prompt disambiguation and retrieval-augmented prompt enhancement were enabled to isolate the impact of error feedback-driven refinement. The evaluation focused on whether integrating runtime error messages into the prompt could improve the executability and correctness of LLM-generated scripts without any human intervention.

The process began with generating a Python script from a simple prompt. If the script executed successfully, it was deemed complete. However, if a runtime error occurred, the corresponding error message was automatically captured and appended to the original prompt. This updated prompt was then fed back into the same LLM, which was instructed to revise the code based on the reported issues. Each new script was re-executed in a controlled environment, and the process repeated iteratively until either the script ran without errors or a maximum of six iterations was reached.

\begin{figure}[h] \centering
\vspace{-2mm}
\includegraphics[width=0.99\columnwidth]{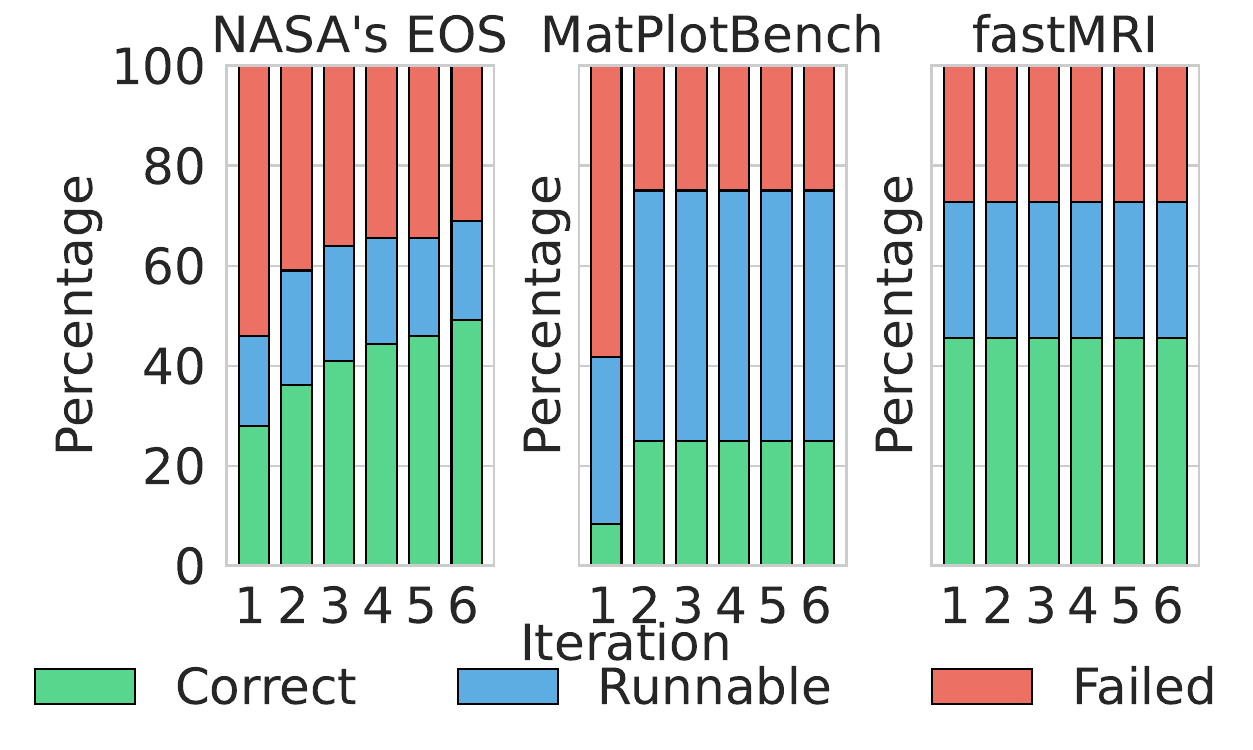}
\vspace{-2mm}
\caption{Execution results of data analysis/visualization codes generated by Devstral-24B using simple prompts (both data-aware prompt disambiguation and retrieval-augmented prompt enhancement are enabled by default) with different number of iterations.}
\label{simple_query_combined_3_datasets_iterative_error_resolve}
\vspace{-2mm}
\end{figure}

As shown in Figure~\ref{simple_query_combined_3_datasets_iterative_error_resolve}, the effectiveness of iterative error repair varied significantly across datasets. For the NASA EOS datasets, code quality steadily improved with each iteration. Many initial errors stemmed from incorrect dataset path usage or incompatible operations on hierarchical data structures. As these errors were surfaced and incorporated into subsequent prompts, the model was able to correct them in a stepwise manner, leading to a noticeable increase in the number of scripts that both executed successfully and produced valid outputs.

In contrast, for the MatPlotBench datasets, improvements were largely front-loaded: the majority of gains occurred within the first iteration, with diminishing returns in subsequent cycles. This behavior suggests that the types of errors encountered in these simpler, tabular datasets were relatively easy to resolve, requiring only one round of correction in most cases. On the other hand, for the fastMRI datasets, the repair mechanism yielded little to no improvement. Many of the scripts that failed in this domain encountered deeply embedded logic or domain-specific processing issues that were not easily resolved through surface-level error message interpretation, even over multiple refinement rounds.

These findings indicate that iterative error repair can be a valuable technique for improving LLM-generated code, especially for tasks involving structured datasets where execution failures stem from syntactic or API-level issues. However, its effectiveness is closely tied to the nature of the dataset and the complexity of the errors.

%% file: 5.related_work.tex
\section{Related Work}
Improving the reliability of LLM-generated code has been the focus of several recent studies, employing techniques such as iterative refinement, retrieval augmentation, ambiguity resolution, and syntax correction. PlotGen~\cite{goswami2025plotgen} refines scientific visualizations using multimodal feedback in a multi-agent framework but is limited to graphical outputs and does not address broader computational workflows involving complex hierarchical data. Similarly,~\cite{DBLP:journals/corr/abs-2311-09868} introduces an interactive LLM-based debugging loop effective for syntax and compilation errors but overlooks structural issues like multi-dimensional indexing or incorrect dataset access—challenges central to scientific data analysis.

Retrieval-augmented generation methods such as REDCODER~\cite{parvez2021retrieval} and CodeRAG~\cite{li2025coderag} improve output quality by incorporating relevant code snippets or graph-structured representations. These approaches depend on curated external repositories or training-time integration. Our use of retrieval augmentation differs in that we dynamically append relevant examples based on parsed user intent, enabling prompt enhancement that adapts to data-specific analysis tasks without requiring external corpora.

Addressing ambiguity in user intent has also been explored through interactive clarification~\cite{10.1145/3660810} and structured prompting strategies~\cite{tony2025promptingtechniquessecurecode}. While effective, such approaches often rely on user interaction or general-purpose heuristics. In contrast, our data-aware prompt disambiguation automatically aligns user queries with dataset metadata, reducing ambiguity without human involvement and tailoring refinement to the structure of scientific data files.

Other work focuses on syntax correction and iterative repair. SynCode~\cite{ugare2024syncodellmgenerationgrammar} enforces grammatical constraints but lacks semantic validation, while AlphaCode~\cite{ridnik2024codegenerationalphacodiumprompt} uses AI-generated test cases to iteratively improve competitive programming submissions. Our study adopts a similar iterative philosophy, applying execution-driven feedback to improve code correctness in scientific contexts where logical errors often stem from misunderstanding data structure rather than algorithmic flaws.

Broader evaluations of LLMs in scientific applications~\cite{jiang2024surveylargelanguagemodels, nejjar2025llms, hong2024data} highlight both potential and limitations but tend to focus on narrow tasks. For example, MatPlotAgent~\cite{yang2024matplotagentmethodevaluationllmbased} benchmarks LLMs for visualization, without addressing data preprocessing or access. Our work extends this line of inquiry by evaluating how LLMs perform across the full analysis pipeline, including data access, transformation, and visualization, on complex structured formats like HDF5. While frameworks such as CodeAgent~\cite{zhang2024codeagentenhancingcodegeneration} and WikiLLaVA~\cite{caffagni2024wikillavahierarchicalretrievalaugmentedgeneration} leverage external knowledge for improved generation, our approach emphasizes self-contained refinement through metadata extraction and example-guided prompting.

In summary, while prior work has introduced promising techniques for improving LLM-based code generation, our study focuses on evaluating their application to scientific data analysis and visualization workflows. By systematically integrating data-aware disambiguation, retrieval-based enhancement, and execution-guided refinement, we assess how these methods perform in addressing structural and semantic challenges unique to scientific data analysis.

%% file: 6.conclusion.tex
\section{Conclusion}
This study systematically evaluates the trustworthiness of LLM-generated code for scientific data analysis and visualization, focusing on challenges posed by structured, hierarchical datasets common in scientific domains. We show that prompt quality significantly affects code correctness and that open-source LLMs often struggle to produce reliable scripts without added guidance. To address these issues, we explore three complementary strategies—data-aware prompt disambiguation, retrieval-augmented prompt enhancement, and iterative error repair—each improving different aspects of code generation. Our findings highlight both the promise and limits of current LLMs in automating scientific computing tasks and suggest a path forward through better prompt engineering and context integration. We release our benchmark and evaluation framework to support future work in building more robust, accessible, and trustworthy LLM-assisted research tools.

%% file: references.bib
@article{DBLP:journals/corr/abs-2311-09868,
  publtype={informal},
  author={Hanbin Wang and Zhenghao Liu and Shuo Wang and Ganqu Cui and Ning Ding and Zhiyuan Liu and Ge Yu},
  title={INTERVENOR: Prompt the Coding Ability of Large Language Models with the Interactive Chain of Repairing},
  year={2023},
  cdate={1672531200000},
  journal={CoRR},
  volume={abs/2311.09868},
  url={https://doi.org/10.48550/arXiv.2311.09868}
}

@article{10.1145/3660810,
author = {Mu, Fangwen and Shi, Lin and Wang, Song and Yu, Zhuohao and Zhang, Binquan and Wang, ChenXue and Liu, Shichao and Wang, Qing},
title = {ClarifyGPT: A Framework for Enhancing LLM-Based Code Generation via Requirements Clarification},
year = {2024},
issue_date = {July 2024},
publisher = {Association for Computing Machinery},
address = {New York, NY, USA},
volume = {1},
number = {FSE},
url = {https://doi.org/10.1145/3660810},
doi = {10.1145/3660810},
journal = {Proc. ACM Softw. Eng.},
month = jul,
articleno = {103},
numpages = {23}
}

@misc{ugare2024syncodellmgenerationgrammar,
      title={SynCode: LLM Generation with Grammar Augmentation}, 
      author={Shubham Ugare and Tarun Suresh and Hangoo Kang and Sasa Misailovic and Gagandeep Singh},
      year={2024},
      eprint={2403.01632},
      archivePrefix={arXiv},
      primaryClass={cs.LG},
      url={https://arxiv.org/abs/2403.01632}, 
}

@misc{jiang2024surveylargelanguagemodels,
      title={A Survey on Large Language Models for Code Generation}, 
      author={Juyong Jiang and Fan Wang and Jiasi Shen and Sungju Kim and Sunghun Kim},
      year={2024},
      eprint={2406.00515},
      archivePrefix={arXiv},
      primaryClass={cs.CL},
      url={https://arxiv.org/abs/2406.00515}, 
}

@misc{yang2024matplotagentmethodevaluationllmbased,
      title={MatPlotAgent: Method and Evaluation for LLM-Based Agentic Scientific Data Visualization}, 
      author={Zhiyu Yang and Zihan Zhou and Shuo Wang and Xin Cong and Xu Han and Yukun Yan and Zhenghao Liu and Zhixing Tan and Pengyuan Liu and Dong Yu and Zhiyuan Liu and Xiaodong Shi and Maosong Sun},
      year={2024},
      eprint={2402.11453},
      archivePrefix={arXiv},
      primaryClass={cs.CL},
      url={https://arxiv.org/abs/2402.11453}, 
}

@misc{tony2025promptingtechniquessecurecode,
      title={Prompting Techniques for Secure Code Generation: A Systematic Investigation}, 
      author={Catherine Tony and Nicolás E. Díaz Ferreyra and Markus Mutas and Salem Dhiff and Riccardo Scandariato},
      year={2025},
      eprint={2407.07064},
      archivePrefix={arXiv},
      primaryClass={cs.SE},
      url={https://arxiv.org/abs/2407.07064}, 
}

@misc{sarker2024syntacticrobustnessllmbasedcode,
      title={Syntactic Robustness for LLM-based Code Generation}, 
      author={Laboni Sarker and Mara Downing and Achintya Desai and Tevfik Bultan},
      year={2024},
      eprint={2404.01535},
      archivePrefix={arXiv},
      primaryClass={cs.SE},
      url={https://arxiv.org/abs/2404.01535}, 
}

@misc{ridnik2024codegenerationalphacodiumprompt,
      title={Code Generation with AlphaCodium: From Prompt Engineering to Flow Engineering}, 
      author={Tal Ridnik and Dedy Kredo and Itamar Friedman},
      year={2024},
      eprint={2401.08500},
      archivePrefix={arXiv},
      primaryClass={cs.LG},
      url={https://arxiv.org/abs/2401.08500}, 
}

@inproceedings{10.1145/1966895.1966900,
author = {Folk, Mike and Heber, Gerd and Koziol, Quincey and Pourmal, Elena and Robinson, Dana},
title = {An overview of the HDF5 technology suite and its applications},
year = {2011},
isbn = {9781450306140},
publisher = {Association for Computing Machinery},
address = {New York, NY, USA},
url = {https://doi.org/10.1145/1966895.1966900},
doi = {10.1145/1966895.1966900},
booktitle = {Proceedings of the EDBT/ICDT 2011 Workshop on Array Databases},
pages = {36–47},
numpages = {12},
location = {Uppsala, Sweden},
series = {AD '11}
}

@misc{zhang2024codeagentenhancingcodegeneration,
      title={CodeAgent: Enhancing Code Generation with Tool-Integrated Agent Systems for Real-World Repo-level Coding Challenges}, 
      author={Kechi Zhang and Jia Li and Ge Li and Xianjie Shi and Zhi Jin},
      year={2024},
      eprint={2401.07339},
      archivePrefix={arXiv},
      primaryClass={cs.SE},
      url={https://arxiv.org/abs/2401.07339}, 
}

@article{achiam2023gpt,
  title={Gpt-4 technical report},
  author={Achiam, Josh and Adler, Steven and Agarwal, Sandhini and Ahmad, Lama and Akkaya, Ilge and Aleman, Florencia Leoni and Almeida, Diogo and Altenschmidt, Janko and Altman, Sam and Anadkat, Shyamal and others},
  journal={arXiv preprint arXiv:2303.08774},
  year={2023}
}

@misc{caffagni2024wikillavahierarchicalretrievalaugmentedgeneration,
      title={Wiki-LLaVA: Hierarchical Retrieval-Augmented Generation for Multimodal LLMs}, 
      author={Davide Caffagni and Federico Cocchi and Nicholas Moratelli and Sara Sarto and Marcella Cornia and Lorenzo Baraldi and Rita Cucchiara},
      year={2024},
      eprint={2404.15406},
      archivePrefix={arXiv},
      primaryClass={cs.CV},
      url={https://arxiv.org/abs/2404.15406}, 
}

@Article{electronics13132657,
AUTHOR = {Bae, Jaehyeon and Kwon, Seoryeong and Myeong, Seunghwan},
TITLE = {Enhancing Software Code Vulnerability Detection Using GPT-4o and Claude-3.5 Sonnet: A Study on Prompt Engineering Techniques},
JOURNAL = {Electronics},
VOLUME = {13},
YEAR = {2024},
NUMBER = {13},
ARTICLE-NUMBER = {2657},
URL = {https://www.mdpi.com/2079-9292/13/13/2657},
ISSN = {2079-9292},
DOI = {10.3390/electronics13132657}
}

@misc{wei2024magicoderempoweringcodegeneration,
      title={Magicoder: Empowering Code Generation with OSS-Instruct}, 
      author={Yuxiang Wei and Zhe Wang and Jiawei Liu and Yifeng Ding and Lingming Zhang},
      year={2024},
      eprint={2312.02120},
      archivePrefix={arXiv},
      primaryClass={cs.CL},
      url={https://arxiv.org/abs/2312.02120}, 
}

@article{guo2025deepseek,
  title={Deepseek-r1: Incentivizing reasoning capability in llms via reinforcement learning},
  author={Guo, Daya and Yang, Dejian and Zhang, Haowei and Song, Junxiao and Zhang, Ruoyu and Xu, Runxin and Zhu, Qihao and Ma, Shirong and Wang, Peiyi and Bi, Xiao and others},
  journal={arXiv preprint arXiv:2501.12948},
  year={2025}
}

@misc{siriwardhana2024domainadaptationllama370binstructcontinual,
      title={Domain Adaptation of Llama3-70B-Instruct through Continual Pre-Training and Model Merging: A Comprehensive Evaluation}, 
      author={Shamane Siriwardhana and Mark McQuade and Thomas Gauthier and Lucas Atkins and Fernando Fernandes Neto and Luke Meyers and Anneketh Vij and Tyler Odenthal and Charles Goddard and Mary MacCarthy and Jacob Solawetz},
      year={2024},
      eprint={2406.14971},
      archivePrefix={arXiv},
      primaryClass={cs.CL},
      url={https://arxiv.org/abs/2406.14971}, 
}

@INPROCEEDINGS{773470,
  author={Ullman, R.E.},
  booktitle={IEEE 1999 International Geoscience and Remote Sensing Symposium. IGARSS'99 (Cat. No.99CH36293)}, 
  title={HDF-EOS, NASA's standard data product distribution format for the Earth Observing System data information system}, 
  year={1999},
  volume={1},
  number={},
  pages={276-278 vol.1},
  doi={10.1109/IGARSS.1999.773470}}

@ARTICLE{56302,
  author={Rew, R. and Davis, G.},
  journal={IEEE Computer Graphics and Applications}, 
  title={NetCDF: an interface for scientific data access}, 
  year={1990},
  volume={10},
  number={4},
  pages={76-82},
  doi={10.1109/38.56302}}

@article{pence2010definition,
  title={Definition of the flexible image transport system (fits), version 3.0},
  author={Pence, William D and Chiappetti, Lucio and Page, Clive G and Shaw, Richard A and Stobie, Elizabeth},
  journal={Astronomy \& Astrophysics},
  volume={524},
  pages={A42},
  year={2010},
  publisher={EDP Sciences}
}

@article{kehrer2012visualization,
  title={Visualization and visual analysis of multifaceted scientific data: A survey},
  author={Kehrer, Johannes and Hauser, Helwig},
  journal={IEEE transactions on visualization and computer graphics},
  volume={19},
  number={3},
  pages={495--513},
  year={2012},
  publisher={IEEE}
}

@inproceedings{fuchs2009visualization,
  title={Visualization of multi-variate scientific data},
  author={Fuchs, Raphael and Hauser, Helwig},
  booktitle={Computer Graphics Forum},
  volume={28},
  number={6},
  pages={1670--1690},
  year={2009},
  organization={Wiley Online Library}
}

@article{steed2013big,
  title={Big data visual analytics for exploratory earth system simulation analysis},
  author={Steed, Chad A and Ricciuto, Daniel M and Shipman, Galen and Smith, Brian and Thornton, Peter E and Wang, Dali and Shi, Xiaoying and Williams, Dean N},
  journal={Computers \& Geosciences},
  volume={61},
  pages={71--82},
  year={2013},
  publisher={Elsevier}
}

@INPROCEEDINGS{10449667,
  author={Fan, Angela and Gokkaya, Beliz and Harman, Mark and Lyubarskiy, Mitya and Sengupta, Shubho and Yoo, Shin and Zhang, Jie M.},
  booktitle={2023 IEEE/ACM International Conference on Software Engineering: Future of Software Engineering (ICSE-FoSE)}, 
  title={Large Language Models for Software Engineering: Survey and Open Problems}, 
  year={2023},
  volume={},
  number={},
  pages={31-53},
  doi={10.1109/ICSE-FoSE59343.2023.00008}}

@article{hong2024data,
  title={Data interpreter: An llm agent for data science},
  author={Hong, Sirui and Lin, Yizhang and Liu, Bang and Liu, Bangbang and Wu, Binhao and Zhang, Ceyao and Wei, Chenxing and Li, Danyang and Chen, Jiaqi and Zhang, Jiayi and others},
  journal={arXiv preprint arXiv:2402.18679},
  year={2024}
}

@article{nejjar2025llms,
  title={Llms for science: Usage for code generation and data analysis},
  author={Nejjar, Mohamed and Zacharias, Luca and Stiehle, Fabian and Weber, Ingo},
  journal={Journal of Software: Evolution and Process},
  volume={37},
  number={1},
  pages={e2723},
  year={2025},
  publisher={Wiley Online Library}
}

@article{gray2005scientific,
  title={Scientific data management in the coming decade},
  author={Gray, Jim and Liu, David T and Nieto-Santisteban, Maria and Szalay, Alex and DeWitt, David J and Heber, Gerd},
  journal={Acm Sigmod Record},
  volume={34},
  number={4},
  pages={34--41},
  year={2005},
  publisher={ACM New York, NY, USA}
}

@inproceedings{springmeyer1992characterization,
  title={A characterization of the scientific data analysis process},
  author={Springmeyer, Rebecca R and Blattner, Meera M and Max, Nelson L},
  booktitle={IEEE Visualization},
  volume={92},
  pages={235--242},
  year={1992}
}

@article{husein2024large,
  title={Large language models for code completion: A systematic literature review},
  author={Husein, Rasha Ahmad and Aburajouh, Hala and Catal, Cagatay},
  journal={Computer Standards \& Interfaces},
  pages={103917},
  year={2024},
  publisher={Elsevier}
}

@article{tian2024debugbench,
  title={Debugbench: Evaluating debugging capability of large language models},
  author={Tian, Runchu and Ye, Yining and Qin, Yujia and Cong, Xin and Lin, Yankai and Pan, Yinxu and Wu, Yesai and Hui, Haotian and Liu, Weichuan and Liu, Zhiyuan and others},
  journal={arXiv preprint arXiv:2401.04621},
  year={2024}
}

@inproceedings{gao2024search,
  title={Search-based llms for code optimization},
  author={Gao, Shuzheng and Gao, Cuiyun and Gu, Wenchao and Lyu, Michael},
  booktitle={2025 IEEE/ACM 47th International Conference on Software Engineering (ICSE)},
  pages={254--266},
  year={2024},
  organization={IEEE Computer Society}
}

@article{sun2024source,
  title={Source code summarization in the era of large language models},
  author={Sun, Weisong and Miao, Yun and Li, Yuekang and Zhang, Hongyu and Fang, Chunrong and Liu, Yi and Deng, Gelei and Liu, Yang and Chen, Zhenyu},
  journal={arXiv preprint arXiv:2407.07959},
  year={2024}
}

@Book{vtkBook,
  author    = "Will Schroeder and Ken Martin and Bill Lorensen",
  title     = "The Visualization Toolkit (4th ed.)",
  publisher = "Kitware",
  year      = "2006",
  isbn      = "978-1-930934-19-1",
}

@article{goswami2025plotgen,
  title={PlotGen: Multi-Agent LLM-based Scientific Data Visualization via Multimodal Feedback},
  author={Goswami, Kanika and Mathur, Puneet and Rossi, Ryan and Dernoncourt, Franck},
  journal={arXiv preprint arXiv:2502.00988},
  year={2025}
}

@misc{hdf_eos_tools_and_information_center,
  author       = {HDF-EOS Tools and Information Center},
  title        = {NASA HDF-EOS5/HDF5 Datapool},
  year         = {2021},
  month        = sep,
  note         = {We collected sample HDF5 and HDF-EOS5 files from NASA data centers and documented information that may be useful.},
  howpublished = {\url{https://hdfeos.org/zoo/Data_Collection/hdf5index.php}}
}

@misc{zbontar2019fastmriopendatasetbenchmarks,
      title={fastMRI: An Open Dataset and Benchmarks for Accelerated MRI}, 
      author={Jure Zbontar and Florian Knoll and Anuroop Sriram and Tullie Murrell and Zhengnan Huang and Matthew J. Muckley and Aaron Defazio and Ruben Stern and Patricia Johnson and Mary Bruno and Marc Parente and Krzysztof J. Geras and Joe Katsnelson and Hersh Chandarana and Zizhao Zhang and Michal Drozdzal and Adriana Romero and Michael Rabbat and Pascal Vincent and Nafissa Yakubova and James Pinkerton and Duo Wang and Erich Owens and C. Lawrence Zitnick and Michael P. Recht and Daniel K. Sodickson and Yvonne W. Lui},
      year={2019},
      eprint={1811.08839},
      archivePrefix={arXiv},
      primaryClass={cs.CV},
      url={https://arxiv.org/abs/1811.08839}, 
}

@article{parvez2021retrieval,
  title={Retrieval augmented code generation and summarization},
  author={Parvez, Md Rizwan and Ahmad, Wasi Uddin and Chakraborty, Saikat and Ray, Baishakhi and Chang, Kai-Wei},
  journal={arXiv preprint arXiv:2108.11601},
  year={2021}
}

@article{li2025coderag,
  title={Coderag: Supportive code retrieval on bigraph for real-world code generation},
  author={Li, Jia and Shi, Xianjie and Zhang, Kechi and Li, Lei and Li, Ge and Tao, Zhengwei and Liu, Fang and Tao, Chongyang and Jin, Zhi},
  journal={arXiv preprint arXiv:2504.10046},
  year={2025}
}

@article{reimers2019sentence,
  title={Sentence-bert: Sentence embeddings using siamese bert-networks},
  author={Reimers, Nils and Gurevych, Iryna},
  journal={arXiv preprint arXiv:1908.10084},
  year={2019}
}

@article{douze2024faiss,
  title={The faiss library},
  author={Douze, Matthijs and Guzhva, Alexandr and Deng, Chengqi and Johnson, Jeff and Szilvasy, Gergely and Mazar{\'e}, Pierre-Emmanuel and Lomeli, Maria and Hosseini, Lucas and J{\'e}gou, Herv{\'e}},
  journal={arXiv preprint arXiv:2401.08281},
  year={2024}
}

@software{ollama2023,
  author = {{Ollama Developers}},
  title = {{Ollama: Run Large Language Models Locally}},
  url = {https://ollama.com/},
  version = {0.7.1},
  year = {2023}, 
  urldate = {2024-08-10} 
}

@misc{Devstral2025,
  author = {{Mistral AI} and {All Hands AI}},
  title = {{Devstral}: An Open-Source Agentic {LLM} for Software Engineering},
  year = {2025},
  url = {https://mistral.ai/news/devstral},
  note = {Available from: \url{https://ollama.com/library/devstral:24b}},   
}

@article{team2025gemma,
  title={Gemma 3 technical report},
  author={Team, Gemma and Kamath, Aishwarya and Ferret, Johan and Pathak, Shreya and Vieillard, Nino and Merhej, Ramona and Perrin, Sarah and Matejovicova, Tatiana and Ram{\'e}, Alexandre and Rivi{\`e}re, Morgane and others},
  journal={arXiv preprint arXiv:2503.19786},
  year={2025}
}

@misc{Neural_Data_Science_in_Python,
  author       = {Aaron J Newman},
  title        = {Neural Data Science in Python},
  year         = {2020},
  note         = {Based on the provided MRI brain image example, several Python scripts have been created to process and analyze data converted from DICOM to HDF5 format. These scripts perform tasks such as viewing metadata, exploring image dimensions and contents, visualizing slices across various planes, rotating images, plotting volume slices, and generating histograms—enabling comprehensive inspection of brain volume data.},
  howpublished = {\url{https://neuraldatascience.io/8-mri/read_viz.html}}
}
